\documentclass[traditabstract]{aa}   

\usepackage{graphicx}
\usepackage{natbib}
\usepackage{txfonts}
\newcommand{\mincir}{\raise
-3.truept\hbox{\rlap{\hbox{$\sim$}}\raise4.truept\hbox{$<$}\ }}
\newcommand{\magcir}{\raise
-3.truept\hbox{\rlap{\hbox{$\sim$}}\raise4.truept\hbox{$>$}\ }}

\begin{document}

\title{Comparison of the spatial and the angular
clustering of X-ray AGN}

   \subtitle{}
   
    \titlerunning{Clustering of X-ray AGN}
    \authorrunning{L. Koutoulidis  et al.}
   
   \author{L. Koutoulidis
          \inst{1,2}
           M. Plionis
          \inst{3,1,4}
 \and
          I. Georgantopoulos
          \inst{1}
          A. Georgakakis
          \inst{5,1}
          A. Akylas
          \inst{1}
          S. Basilakos
          \inst{6}
          G. Mountrichas
          \inst{1}
          }
     \institute{IAASARS, National Observatory of Athens, I. Metaxa \& V. Pavlou 1, Penteli, 15236, Greece \\
     \and
    Astronomical Laboratory, Department of Physics, University of Patras, 26500 Rio-Patras, Greece \\
    \and
   Physics Department, University of Thessaloniki, Thessaloniki 54124, Greece \\
   \and
      Instituto Nacional de Astrof\'isica, \'Optica y Electronica, Luis Enrique Erro l, Tonantzilla, Puebla, M\'exico \\
      \and
    Max-Planck-Institut f$\ddot{u}$r Extraterrestrische Physik (MPE),
    Postfach 1312, 85741, Garching, Germany \\
     \and
    Academy of Athens, Research Center for Astronomy and Applied Mathematics, Soranou Efesiou 4, 11527, Athens, Greece}                           
   \date{}

 
\abstract
{The angular correlation function is a powerful tool for deriving the 
clustering properties of AGN and hence the mass of the corresponding dark matter halos 
in which they reside. However, recent studies based on the application 
of the angular correlation function on X-ray samples, yield results apparently 
inconsistent  with those based on the direct estimation of the spatial 
correlation function. The goal of the present paper is to attempt to
investigate this issue by analysing a well defined sample. 
 To this end we use the hard-band (2-10 keV)
  X-ray selected sources of the Chandra AEGIS fields, 
chosen because of the availability of accurately derived flux sensitivity maps.
In particular we use the 186 hard-band sources with
spectroscopic redshifts in the range $z=0.3-1.3$,
a range selected in order to contain
the bulk of the AGN while minimizing the contribution of unknown
clustering and luminosity evolution from very high redshifts. 
Using the projected spatial auto-correlation function, we derive a
 clustering comoving length of $x_0=5.4 \pm 1.0 \;h^{-1}$ Mpc (for
 $\gamma=1.8$), consistent with results in the  literature.
We further derive the angular correlation function and the corresponding 
spatial clustering length using the Limber's inversion equation 
and a novel parametrization of the clustering evolution model that
also takes into account the bias evolution of the host dark matter
halo. The Limber's inverted spatial comoving clustering length of
$x_0=5.5\pm 1.2 \;h^{-1}$
Mpc at a median redshift of $z\simeq 0.75$, matches the directly
measured one, from the spatial correlation function analysis, 
but for a significant non-linear
contribution to the growing mode of perturbations, 
estimated independently from literature results of $x_0$ at different redshifts. 
Therefore, using this sample of hard X-ray AGN 
and our clustering evolution parametrization
 we have found an excellent consistency between the
angular and spatial clustering analysis.
 }

 \keywords{active galaxies --
               clustering --
                X-rays--
                galaxies               }
  \authorrunning{L. Koutoulidis et al.}

\maketitle

\section{Introduction}\label{sec_intro}
The triggering mechanism for the AGN activity is still an open issue
\citep{Alexander12}. 
Several semi-analytical models which 
include  major galaxy mergers can explain the triggering mechanism for
the most luminous AGN \citep{DiMatteo05,Hopkins06} 
while it is possible that in the lowest luminosity AGN regime secular
evolution (disk instabilities or minor interactions) 
may play the key role \citep{HopkinsHernquist06,Bournaud11}.

Measuring the clustering of AGN can place valuable constraints on the
AGN fueling modes and also provide us important 
information for the AGN activity and their dark matter halo
hosts. Merger models appear to reproduce the clustering of 
QSOs and the mass of dark matter halo in which they reside. However in
the X-ray regime, clustering of X-ray AGN shows 
evidence that they live in more massive dark matter halos, one order
of magnitude larger \citep{Koutoulidis13} than the 
optical QSOs. This result suggest that the main accretion mode is the
so-called hot halo mode \citep{Fanidakis12,Fanidakis13}. 

The clustering of AGN has been studied with excellent number
statistics in the optical bands, particularly in large area surveys
such as the 2QZ \citep{Croom05, Porciani06} and the SDSS (Sloan
Digital Sky Survey) \citep{Li06, Shen09, Ross09}. However, optical QSO
may represent only the tip of the iceberg of the AGN population. 
X-ray surveys find a surface density of about
20,000 $\rm{deg^{-2}}$ \citep{Xue11} which is about two orders of
magnitude higher than that found in optical QSO surveys \citep{Wolf03}. 
X-ray selected, spectroscopically identified AGN, form a superset of
the optical selected AGN population since a large fraction of hard
X-ray selected AGNs do not show strong optical activity \citep{Barger2005}.
Therefore, in order to study the clustering of the total AGN
population we need X-ray samples.
Recently, several studies have attempted to measure the spatial
 correlation function of X-ray selected AGN, using spectroscopic redshifts
 to estimate their distances at moderate redshifts \citep{Mulli04,
   Gilli05, Yang06, Gilli09, Hickox09, Coil09, Krumpe10, 
 Starikova11, Allevato11, Koutoulidis13} and at low redshifts
 \citep{Cappelluti10}. Better statistics 
can be achieved using a cross correlation analysis with galaxies,
either using spectroscopic AGN and galaxy samples
\citep{Coil09,Krumpe10,Mountrichas2012} or only spectroscopy for the
AGNs and photometric redshifts (their pdfs) for the galaxies
\citep{Mountrichas2013}. In order to derive directly the spatial
clustering length for a large sample of X-ray AGN an extensive
spectroscopic campaign is required or high quality of photometric
redshift measurements. However, even better statistics can be provided
by the angular correlation function (ACF) for which all the detected
sources are used, independently of the availability of spectroscopy.

Several studies have explored the angular clustering of AGN  in X-ray
wavelengths  using data from {\it ROSAT} \citep{Vikhlinin95,Akylas2000}, 
from {\it XMM-Newton} \citep{Basilakos04,Basilakos05,Puccetti06,Ebrero09, 
Miyaji07,Elyiv12} or deep pencil CDFs fields \citep{Gilli05,Plionis08}.
These studies measure the projected
angular clustering and then via Limber's equation \citep{Peebles1980} 
derive the corresponding spatial clustering length. 
Their results however appear to contradict the direct measurements of spatial 
clustering, with all the angular correlation analyses finding systematically larger 
correlation amplitudes. Possible reasons for this discrepancy include
uncertainties in the X-ray AGN luminosity function
and thus in the corresponding redshift distribution as well as the the
clustering evolution model, both of which 
are necessary for the Limber's inversion.

A way to break this impasse is to derive both the angular and spatial
clustering for the same set of objects and compare directly their results.
In this paper we derive the spatial correlation function in the AEGIS
field in the hard band, using 186 sources with 
spectroscopic redshift information. Then we derive the angular
correlation function (ACF) for exactly the same sources, to 
infer the spatial correlation length, using in one case the 
redshift distribution providing from luminosity function and 
in the other case the redshift distribution as it observed form the
sources with spectroscopic redshifts. 

The paper is organised as follows. In section 2 we present the AEGIS
data, in section 3 we present our methodology and 
our modeling of the clustering evolution, while the results from 
our spatial and angular correlation function analysis, the
Limber's inversion of the latter and the comparison of the two
clustering lengths are presented in section 4.
Throughout this work we adopt a flat $\Lambda$CDM cosmological model 
with $H_0=100$ kms$^{-1}$ Mpc$^{-1}$.

\section{AEGIS CATALOG}
The ultra deep field survey comprises of pointings at 8 separate
positions, each with a nominal exposure 200 ksec, covering a total
area of approximately 0.67 deg$^2$ and centered at $a=14^h 17^m$, 
$\delta=+52^{\circ} 30^{'}$ in a strip of length 2 degrees with a flux
limit of $3.8 \times 10^{-16}$ erg s$^{-1}$cm$^{-2}$ in the hard band. We use
X-ray source catalogue of \citet{Laird09}.  In the hard band we have a
total of 741 X-ray sources. Spectroscopic redshifts are available from
the DEEP2 survey \citep{Davis01,Davis03,Coil09} for 312 sources spanning the 
$0<z<4.3$ range. However, in the current work we will restrict our
analysis to within the redshift interval $z=0.3-1.3$,
comprising 186 sources, in order to minimize strong
evolutionary effects of the spatial correlation function and of
the hard-band X-ray luminosity 
function, while having enough sources to obtain a relatively
robust clustering signal. 
The median redshift of this spectroscopic subsample is $\bar{z}=0.75$.

The advantage of using the AEGIS field for the purpose of this study 
is that source detection and sensitivity maps are constructed
self-consistently following the method described in detail in
\citet{Laird09,Georgakakis08}, which is an analytical method
that accurately estimates the probability of detecting a source
with a given X-ray flux at a given position on the detector accounting
for vignetting and flux biases.
Such sensitivity maps are essential for the production of accurate random 
catalogues and therefore for the reliable determination of the angular
correlation function. 
It is likely that inconsistencies between the spatial and angular correlation function
could, at least partially, originate from using sensitivity maps which are not consistent
with the provided X-ray source lists. 

 \section{METHODOLOGY}
In order to quantify the low-order clustering of a distribution of sources, 
one uses the two-point correlation function which describes the excess
probability over random of finding pairs of
sources within a range of separations.

Depending on the availability or not of redshifts, one can use the 
spatial or angular correlation function. The former, $\xi(r)$  
involves sources within elemental volumes $dV_i$
separated by a distance $r$ \citep[e.g.,][]{Peebles1980}, and is given by: 
$dP = \langle n\rangle^2 [1+ \xi(r)] dV_1dV_2$, where $\langle n\rangle$ 
is the mean space source density. 

If redshifts are not available, one can measure the angular correlation 
function, $w(\theta)$, on the plane of sky which involves finding pairs of 
sources within infinitesimal 
solid angles, $d\omega_i$, separated by an angle $\theta$.
The equivalent mathematical description is given by 
$dP = \langle n\rangle^2 [1+ w(\theta)] d\omega_1 d\omega_2$. 
On small scales $w(\theta)$ has also been found to follow a power 
law behaviour: $w(\theta)=(\theta/\theta_0)^\beta$ with $\beta\simeq 1-\gamma$.

The actual correlation function estimator used, being either spatial
or angular (generically indicated with ${\cal W}$),
is given by the expression \citep{Hamilton93}
\begin{equation}\label{LS}
{\cal W}={\cal N}\frac{DD \times RR}{DR^{2}}-1
\end{equation} 
with $DD$, $RR$ and $DR$ the
data-data, random-random, data-random pairs, respectively, at some
separation $r$ or $\theta$, while ${\cal N}$ is a small correction equal to
the ratio $(N_D N_R)^2/N_D(N_D-1)N_R(N_R-1)$ where $N_D$ and $N_R$ are
the numbers of real and random data, respectively.

The variance of the correlation function at each separation is estimated
according to: 
\begin{equation}\label{error}
\sigma^2_{\cal W}=3 \frac{(1+{\cal W})^2}{DD}\;,
\end{equation}
which corresponds to that expected by the bootstrap resampling technique
\citep{Mo92}.
In order to estimate from the derived correlation function the values
of the correlation length ($r_0$ or $\theta_0$)
and of the slope $\gamma$, we use a $\chi^2$ minimization procedure
between the derived ${\cal W}$ 
and the power law model:
\begin{equation}
\chi^2(r_0,\gamma)=\sum_{i=1}^{n}\frac{({\cal W}_{data}-{\cal
    W}_{model})^2}{\sigma_{\cal W}^2}
\label{eq:chi2}
\end{equation}
where $n$ is the number of separation bins. The minimization is over
scales where the power-law appears to be a reasonable fit to the data
(thus, very large and very small scales are excluded from the fit,
which in
our case translate to: $r_p\lesssim 1$ and $r_p\gtrsim 20 \; h^{-1}$ Mpc).

\subsection{Modelling the 2-point correlation function}
The spatial correlation function of all different mass tracers of the large-scale
structure of the Universe, being galaxies, AGN or clusters of
galaxies, is described well by a power-law with two free parameters,
$r_0$ and $\gamma$, the former related to the amplitude of clustering
and the latter on the slope of the power-law. Locally, at $z\simeq 0$,
it takes the form:
\begin{equation}\label{eq:powerlaw}
\xi(r)=\left(\frac{r}{r_0}\right)^{-\gamma}
\end{equation}
where $r$ is the proper separation between any two tracers. 
It has been found that
$\gamma\simeq 1.8$ for a wide range of mass tracers. In an evolving
Universe the spatial correlation function 
is a function of redshift, as 
expected from the fact that the density perturbations evolve with
redshift. 

Traditionally the evolution of clustering has been parametrized with the
use of the $\epsilon$ parameter, proposed originally within the framework of the 
EdS model \citep{Groth1977,   Peebles1980, deZotti1990}, which characterizes the different 
clustering evolution models according to:
\begin{equation}
\label{xi223}
\xi(r,z)=(1+z)^{-(3+\epsilon)} \xi(r,0) 
\end{equation} 
Evidently, the value $\epsilon=-3$ for which $\xi(r,z)=\xi(r,0)$
corresponds to constant clustering in proper coordinates,
the value $\epsilon=0$ corresponds to the stable
clustering scenario in which clusters remain bound and stable while
due to the the expansion the background density drops by $(1+z)^3$,
the value
$\epsilon=-1$ corresponds to the linear evolution (within
the EdS model and neglecting the evolution of bias), while for
$\epsilon=\gamma-3$ the clustering is constant in comoving
coordinates, as it can be appreciated from the evolution of the 
correlation length given, for the case of a power law 
(eq.\ref{eq:powerlaw}), by:
\begin{equation}
r_0^{\gamma}(z)= \left(\frac{x_0}{1+z}\right)^{\gamma}=r_0^{\gamma}
(1+z)^{-(3+\epsilon)}
\end{equation}
and thus
\begin{equation}\label{tradM}
x_0=r_0 (1+z)^{-(3+\epsilon-\gamma)/\gamma}
\end{equation} 
where $x_0$ is the comoving correlation length at redshift $z$
(and $r_0=x_0(0)$). 
We can now see that indeed if $\epsilon=\gamma-3$ we have a
constant correlation length in comoving coordinates.

In this work we prefer to use a more physical modelling that takes
into account both the cosmological evolution of perturbations and of the linear
bias of the mass tracers. 
We start from the definition of the correlation function, given by:
\begin{equation}
\label{ssep2}
\xi(r,z)=\langle \delta({\bf x},z) \delta({\bf x+r},z)\rangle=
b^{2}(z)\langle \delta_{m}({\bf x},z) \delta_{m}({\bf x+r},z)\rangle
\end{equation}
where
$\delta$ and $\delta_{m}$ are the linear density contrasts 
for the tracers (in our case X-ray AGNs) and for the 
underlying Dark Matter (DM), respectively, and $b(z)$ is the evolution
of the linear bias factor. Therefore, since $\delta_m$ evolves in the
linear regime according to:
\begin{equation}
\delta_m(r,z) = \frac{D(z)}{D(0)} \delta_m(r,0)
\end{equation}
with $D(z)$ the linear growing mode of the density perturbations, we have that:
\begin{equation}\label{eq:xi-r}
\xi(r,z)= \frac{D^2(z)}{D^2(0)} b^2(z) \xi_m(r,0) =
\frac{D^2(z)}{D^2(0)} \frac{b^2(z)}{b^2(0)} \xi(r,0) \;,
\end{equation}
with $\xi(r,0)$ the present epoch spatial correlation function of the X-ray AGN. 
Introducing the normalised to the present linear growing mode and bias as 
${\tilde D}(z)=D(z)/D(0)$ and ${\tilde b}(z)=b(z)/b(0)$, respectively, we write 
Eq.(\ref{eq:xi-r}) as: 
$\xi(r,z)= {\tilde D}^2(z) {\tilde b^2}(z) \xi(r,0)$.
Note however that scales of a few Mpc, ie., around the clustering length of 
the spatial correlation function, should be affected by non-linear 
effects and therefore modelling its evolution only by the linear factor 
${\tilde D}^2(z)$
factor is inadequate. We therefore introduce a further factor ${\tilde D}^n(z)$
where the exponent $n$ absorbs the non-linear effects. Therefore,
Eq.(\ref{eq:xi-r}) becomes:
\begin{equation}\label{eq:xi-r1}
\xi(r,z)= {\tilde D}^{2+n}(z) {\tilde b^2}(z) \xi(r,0)\;,
\end{equation}
and for a power law correlation function we obtain the evolution of the
correlation length in proper coordinates, $r_0(z)$, as:
\begin{equation}
r_0^{\gamma}(z)= \left(\frac{x_0}{1+z}\right)^{\gamma}=r_0^{\gamma}
{\tilde D}^{2+n}(z) {\tilde b^2}(z)
\end{equation}
which translates in comoving coordinates to:
\begin{equation}\label{eq:ourM}
x_0=r_0 (1+z) {\tilde D}^{(2+n)/\gamma}(z) {\tilde b^{2/\gamma}}(z) \;.
\end{equation} 

It is important to appreciate the correspondence of the two different
parametrizations of the clustering evolution. 
Comparing the generic formulation of eq.(\ref{eq:xi-r1}) with that of
the $\epsilon$ paremetrization (eq. \ref{xi223}) 
we see that what is implied is the equivalence of 
${\tilde D}^{2+n}(z) {\tilde b^2}(z)$ with $(1+z)^{-(3+\epsilon)}$,
which however is valid only for as long as both ${\tilde D}(z)$ (as for example in the 
EdS model) and ${\tilde b}(z)$ (as for example in the usual galaxy conserving 
bias model \citep{Fry96}) are power law functions of $1+z$.
However, in the general case of other than the EdS cosmological models
and of a more general bias evolution model, valid also at large
redshifts, \citep[e.g.,][]{Sheth99,Tinker10,Basilakos2008}, the above inferred
equivalence and thus the $\epsilon$ parametrization are not valid.
For example, for the $\Lambda$CDM model the growing mode for the
evolution of linear perturbations is given by \citep[e.g.,][]{Peebles1993}:
\begin{equation}\label{eq:Dz}
D(z)=\frac{5\Omega_{m,0}E(z)}{2}{\int_{z}^{\infty}\frac{(1+y)}{E^3(y)} dy}
\end{equation}
where $E(z)=[\Omega_{m,0}(1+z)^3+\Omega_{\Lambda}]^{1/2}$ 
(for a flat $\Lambda\ne 0$ model), while $\Omega_{m,0}(z)$ and
$\Omega_{\Lambda}(z)$ are respectively the parametrized matter and cosmological
constant density parameters. An EdS look-alike (normalised) fitting
function of eq.(\ref{eq:Dz}) is:
\begin{equation}\label{gr_L}
{\tilde D}(z)=\frac{g(z)}{g(0)}\frac{1}{1+z}=\frac{g(z)}{g(0)}{\tilde D}_{EdS}(z)
\end{equation}
where $g(z)$ is a function of $\Omega_{m,0}(z)$ and
$\Omega_{\Lambda}(z)$ (Carroll, Press \& Turner 1992; Lahav \& Suto 2003).

It will be instructive to graphically compare the two parametrizations and
appreciate their differences. To this end we use 
the bias evolution scheme of  
\cite{Basil01,Basil03} which is based on linear 
perturbation theory and 
given by \citep{Basilpouri11}:
\begin{equation}
b(z)=1+ \frac{b_0 -1}{D(z)}+C_2 \frac{J(z)}{D(z)}
\end{equation}
with
\begin{equation}
J(z)= \int_{0}^{z} \frac{(1+y)}{E(y)}dy \;.
\end{equation}
The constants $b_0$ (the present day bias factor) 
and $C_2$ depend on the host dark matter halo mass and for the
$\Lambda$CDM model are given by:
\begin{equation}
b_0(M_h)=0.857\left[1+\left(C_m \frac{M_h}{10^{14} h^{-1}M_{\odot}}\right)^{0.55}\right]
\end{equation}

\begin{equation}
C_2(M_h)=1.105\left[1+\left(C_m \frac{M_h}{10^{14} h^{-1}M_{\odot}}\right)^{0.255}\right]
\end{equation}
where $C_m=\Omega_{m,0}/0.27$ \citep{Basil12}. This bias evolution
model has been thoroughly tested and found to rate very well in reproducing
N-body simulation data \citep{Basilakos2008} as well as fitting observational data
\citep{Papageo12}.

\begin{figure}\label{models}
\begin{center}
 \includegraphics
[width=8.1cm]{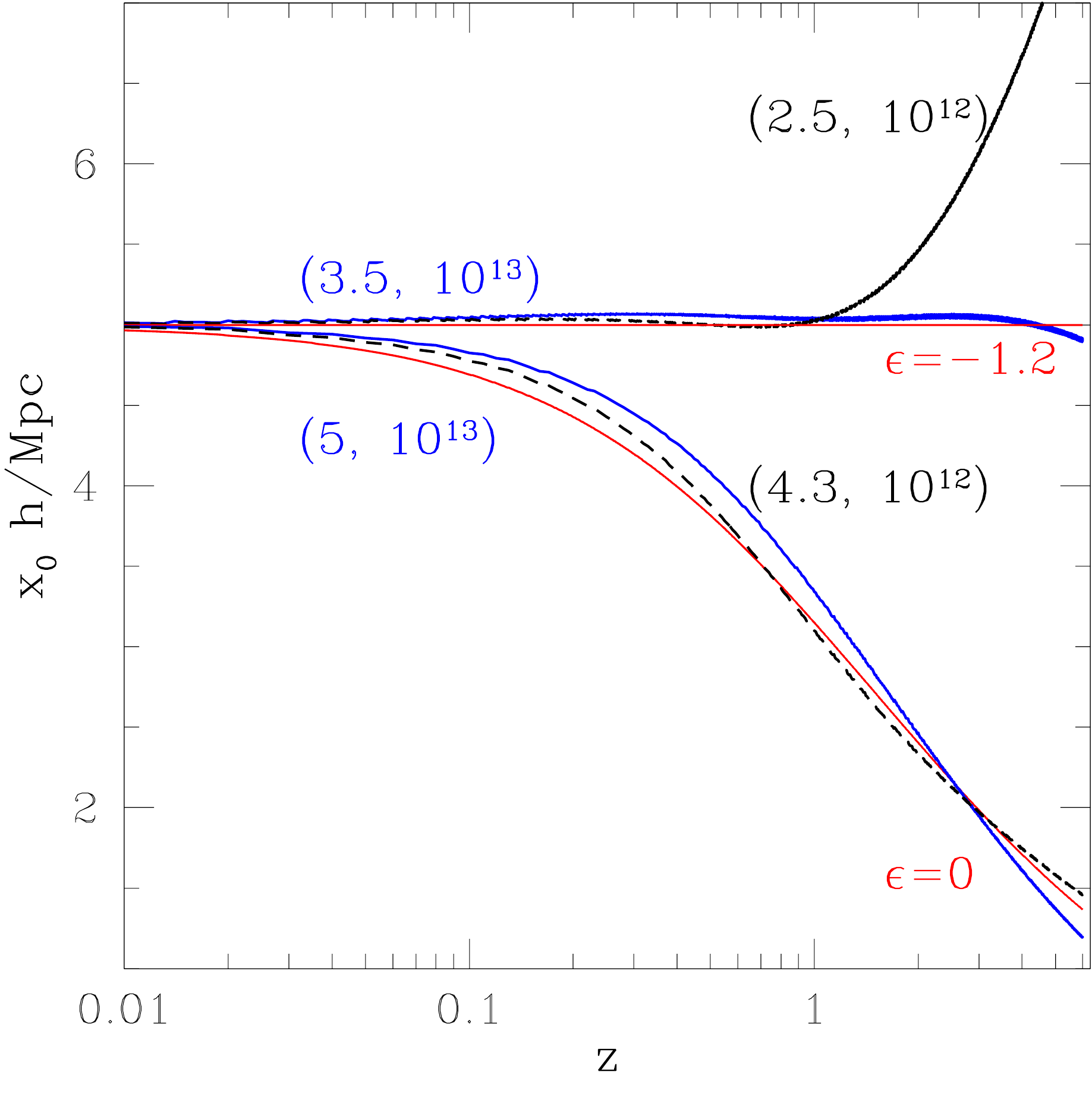} \vfill
\end{center}
\caption{The redshift evolution of the comoving clustering length for the two
  different parametrizations of the
  clustering evolution. For the case of our model (eq.\ref{eq:ourM}),
  shown as black and blue curves, the different
  evolution behaviors shown correspond to the indicated 
parameters $(n, M_h$). The specific values of these parameters 
have been chosen such as to resemble 
the $\epsilon=-1.2$ and $\epsilon=0$ models of the
traditional parametrization of eq.(\ref{tradM}), shown in
red, which correspond to the the constant in comoving coordinates
clustering (for $\gamma=1.8$) and to the stable clustering models (for
details see the text).}
\end{figure}

 We therefore see that through the dependence of the bias factor on
 the dark matter halo mass one expects for the same values of $n$ and
 $\gamma$ a different clustering evolution for dark matter halos of
 different mass. In order to appreciate our parametrization of the
 clustering evolution we present in Figure 1
the evolution of the
 comoving clustering length $x_0$ for models which have
 been on purpose selected to resemble some $\epsilon$ based models.
As can be seen the model with $(n, M_h)=(3.5, 10^{13} M_{\odot})$ is 
equivalent with the constant in comoving coordinates clustering model
($\epsilon=\gamma-3=-1.2$) up to very large redshifts ($z\sim 6$). No other
($n, M_h)$ combination can provide such an equivalence up to such high
redshifts. However, as we can again see in Fig.1 one can find ($n, M_h$) 
combinations
that correspond to the comoving clustering model but up to $z\simeq 1.2$,
as the example shown with $(n, M_h)=(2.5, 10^{12}M_{\odot})$. In fact
one can derive the degeneracy between the $n$ and $M_h$ parameters
such that the clustering evolution is
 constant in comoving coordinates ($\epsilon=-1.2$), but up to
$z\sim 1.2$, since beyond this redshift the equivalence is
unattainable. In Figure 2
we present the corresponding 1, 2 and 3 $\sigma$ contours 
in the $(n, M_h$) parameter space of the 
solution which provides the above
equivalence with the $\epsilon=-1.2$ model. The range is dictated by
the uncertainty which we have imposed on $x_0$, which is taken to be
$10\%$ of $x_0$, typical of current accurate measurements. The degeneracy is clear
and it can only be broken if we either impose a value of $n$, for
example from the expected slope of the power spectrum on these scales,
or from an independent estimate of the mass of the dark matter halos
in which the tracers (in our case X-ray selected AGN) reside.
\begin{figure}\label{comov_cont}
\begin{center}
 \includegraphics
[width=8.1cm]{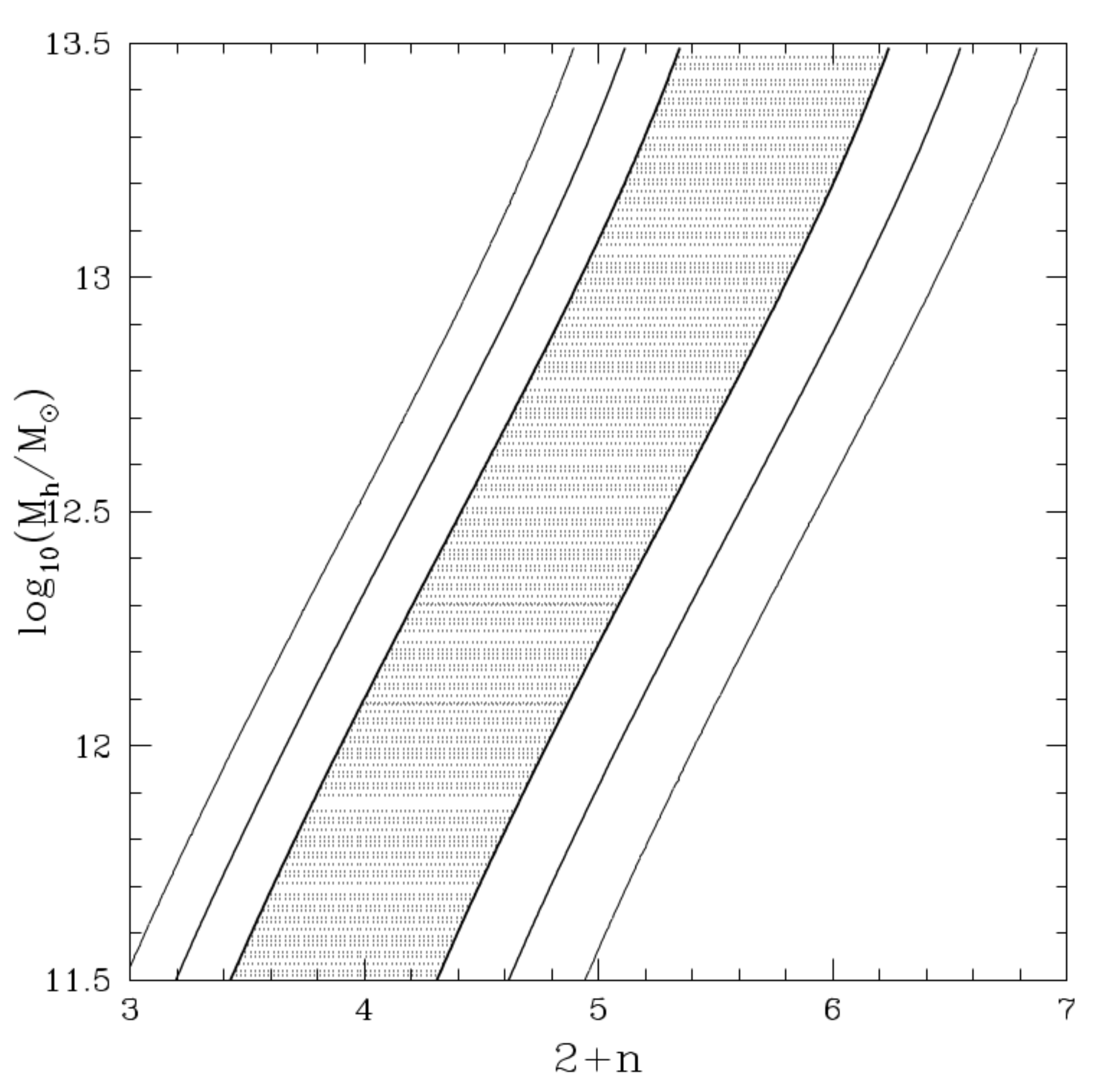} \vfill
\end{center}
\caption{The 1, 2 and 3 $\sigma$ contour range in the $n, M_h$
  parameter plane for the case where our clustering evolution scheme
  corresponds to the constant in comoving coordinates
  model (imposed up to $z=1$ in this example). 
 The strong degeneracy of the parameters is evident.}
\end{figure}

Similarly, larger values of $n$ provide clustering evolution behaviours
that start resembling the $\epsilon=0$ stable clustering model. For
example, we show in Fig.1 the $(n, M_h)=(5, 10^{13}M_{\odot})$ and the
$(4.3, 10^{12}M_{\odot})$ which closely follow the $\epsilon=0$ model,
indicating again the degeneracy problem discussed previously.

In what follows we will use a value for the
dark matter halo mass derived from our previous spatial clustering
analysis of Chandra X-ray selected AGN \citep{Koutoulidis13}, 
ie., $M_h\simeq 1.3 \times 10^{13} h^{-1}M_{\odot}$. As we
already saw, for this value of the halo mass and for $n\simeq 5.5$ one
obtains the constant in comoving coordinates model for the evolution
of clustering.

\subsection{Estimating the Spatial Correlation Function}
Although the recessional velocities of extragalactic sources are used as 
distance indicators, they are contaminated by local peculiar velocities 
and thus the corresponding distances are distorted by the so-called 
redshift-space distortion effect.
An estimator that avoids such effects, while using redshifts to infer distances,
is the projected correlation function $w_p(r_p)$
\citep{DavisPeeb1983}, which is based on deconvolving the
redshift-based comoving distance, $s$, in a component parallel and
perpendicular to the line of sight, $\pi$ and $r_p$, respectively, i.e.,
$s^2=r_p^2+\pi^2$.
The redshift-space correlation function can therefore be written as:
\begin{equation}
\xi(s)=\xi(r_p,\pi)=\xi\left(\sqrt{r_p^2+\pi^2}\right)  \;,
\end{equation}
and the so-called projected correlation function can be found
by integrating $\xi(r_p,\pi)$ along the $\pi$ direction:
\begin{equation}\label{eq:wp}
w_p(r_p)=2\int_{0}^{\infty}\xi(r_p,\pi) \mathrm{d}\pi \;.
\end{equation}
Then the real space correlation function can be recovered
according to \citep{DavisPeeb1983}:
\begin{equation}\label{eq:wp}
 w_p(r_p)=2\int_{0}^{\pi_{\rm max}}\xi\left(\sqrt{r_p^2+\pi^2}\right) 
{\rm d}\pi =2\int_{r_p}^{\infty}
 \frac{x \xi(x)\mathrm{d}x}{\sqrt{x^2-r_p^2}}\;.
 \end{equation}
Modelling $\xi(x)$ as a power law one obtains:
\begin{equation}\label{eq:wp_model}
w_p(r_p)=H_\gamma r_p \left(\frac{x_0}{r_p}\right)^{\gamma}
\end{equation}
with $x_0$ the comoving clustering length at the effective (median) redshift of
the sample, and 
\begin{equation}
H_\gamma=\Gamma\left(\frac{1}{2}\right)
\Gamma\left(\frac{\gamma-1}{2}\right)/\Gamma\left(\frac{\gamma}{2}\right)
\end{equation} 
with $\Gamma$ the usual gamma function.
Note that eq.(\ref{eq:wp}) holds strictly for $\pi_{\rm max}=\infty$,
while in order to avoid redshift-space distortions
the integral is performed up to a finite value of
$\pi_{\rm max}$, which in turn produces an underestimation of the underlying
correlation function. For a power law correlation function this
underestimation is easily
inferred from Eq.(\ref{eq:wp}) and is given by \citep[e.g.,][]{Starikova11}:
\begin{equation}\label{eq:CorF}
C_{\gamma}(r_p)=\frac{\int_0^{\pi_{\rm max}} (r_p^2+\pi^2)^{-\gamma/2} d\pi}
{\int_0^\infty (r_p^2+\pi^2)^{-\gamma/2} d\pi}\;.
\end{equation}
The free of redshift-space distortions correlation function, 
taking into account the above statistical correction 
and under the power-law assumption,
is then provided by:
\begin{equation}\label{eq:corr}
\xi(r_p)=\frac{1}{H_\gamma C_{\gamma}(r_p)} \frac{w_p(r_p)}{r_p} \;.
\end{equation}
which can then be fitted to the power-law model (using
eq.\ref{eq:chi2}) in order to estimate the final and corrected values
of the clustering amplitude and slope.
Alternatively, one can crudely derive the corrected correlation amplitude
(for the slope $\gamma$ and amplitude $x_0$ estimated by fitting
eq.\ref{eq:wp_model}) by: 
\begin{equation}\label{eq:crude}
x_{0,c}\simeq x_0 C_\gamma(x_0)^{-1/\gamma}\;,
\end{equation}
\citep[e.g.,][]{Starikova11,Koutoulidis13}. Note that the correction
factor values range between $C_{\gamma}\simeq 0.85$, at $r_p=3 h^{-1}$ Mpc, and
$C_{\gamma}\simeq 0.45$, at $r_p=10 h^{-1}$ Mpc, (for the $\pi_{max}=10 h^{-1}$ Mpc case).

\subsection{Estimating the Angular Correlation Function and Limber's Inversion}
Another approach that completely avoids redshift-space distortion effects 
is to measure the angular correlation 
function, and then under some assumptions to infer 
the spatial correlation function through the Limber's inversion equation \citep{Limber1953}.

To this end we estimate the angular correlation function in angular logarithmic bins, 
covering the range $10''<\theta< 3800''$, for both the complete source sample and 
the spectroscopic subsample.
Furthermore, we numerically estimated the amplitude of the integral constraint correction 
\citep{Roche1999} which however was found to be negligible, and thus
we neglect it in what follows.

It has been demonstrated \citep{Limber1953} 
that the angular correlation function, $w(\theta)$, can be 
deprojected to yeal the spatial one, $\xi(r)$, via an integral equation.
Under the power law representation of $\xi(r)$ (eq.\ref{eq:powerlaw})
the angular correlation length, $\theta_0$, is related to the corresponding
spatial one, $x_0$, at $z=0$ according to \citep{Peebles1980}:
\begin{equation}\label{eq:inv22}
\theta_0^{\gamma-1} =H_{\gamma}x_0^{\gamma} \int_{0}^{\infty} 
\left(\frac{1}{N}\frac{dN}{dz}\right)^2 
\frac{ {\rm d}_A(z)^{1-\gamma}}{c d\tau(z)/dz} {\tilde b^2}(z) {\tilde D}^{2+n}(z)
dz \;,
\end{equation}
where $\rm{d}_A$ is the angular diameter distance, $\tau(z)$ is the
look-back time and $dN/dz$ is the number of sources per unit redshift
interval within a solid angle $\omega_s$ given by:
\begin{equation}
\frac{\mathrm{d}N}{dz}=\omega_s {\rm d}_A(z)^2 (1+z)^2 \phi(z)\left(\frac{c}{H_0}\right) \frac{1}{E(z)}
\end{equation}
with $\phi(z)=\int_{L_{\rm min(z)}}^{\infty}\Phi(L,z)\mathrm{d}L$ the redshift selection 
function of the sources, i.e., the probability that a source at a comoving distance $x$ 
is detected, and $\Phi(L,z)$ is the luminosity function of the 
sources\footnote{In our case the sources are X-ray  selected AGN and
  the luminosity function that we use is that of the hard-band from
  \citep{Aird2010}.}.

\subsection{Construction of the Random Catalogues}
In order to create the comparison random sample, mimicking the source
catalogue systematic effects and biases, we follow the standard
approach, according to which each
simulated source is placed at a random position on the part of the sky covered by the survey in 
hand, with a flux randomly extracted from the observed source $\log
N-\log S$ \citep{Georgakakis08}. If the flux is above the value allowed by the sensitivity map at that
position, the simulated source is kept in the random sample.
In the current work we use the AEGIS field sensitivity maps of Laird
et al. (2009) 

For the spatial correlation function a random redshift is also assigned to each source from the 
observed source redshift distribution $N(z)$ (optimally taking into
account its variation as a function of flux).
As a test of possible disadvantages of this method, caused by the fact that it does not take
into account any unknown inhomogeneities and systematics of the follow-up
spectroscopic observations, we follow the alternative random catalogue construction approach
of \cite{Gilli05} (hereafter G05). This is based on keeping unaltered
the angular coordinates of the sources while reshuffling their redshifts, 
smoothing the corresponding redshift distribution. For the smoothing we use a Gaussian
kernel with a standard deviation of $\sigma_z=0.2$. This offers a compromise between scales
that are either too small, and thus may reproduce the $z$-space
clustering, or too large and thus over-smooth the observed
redshift distribution. We verified that our results do not change significantly
when using the range: $\sigma_z =0.1-0.3$.  

\begin{table}
\caption{Spatial clustering results for the spectroscopic subsample of the 
AEGIS field (186 sources within $0.3<z<1.3$). The clustering length
units are $h^{-1}$ Mpc.
 The results correspond to $\pi_{\rm max}=10 \; h^{-1}$ Mpc. 
The first row indicates the results based on projected correlation
function $w_p(r_p)$,
while the second corresponds to the corrected spatial correlation
function $\xi(r_p)$ after introducing the correction factor
$C_\gamma(r_p)$ (eq.\ref{eq:corr}). A difference of $\delta x_0 \simeq
+0.9$ is found (for the $\gamma=1.8$ case), corresponding to an
increase of $\sim 20\%$ with respect to the uncorrected correlation
length value.}
\label{tab_wrpsample}

\tabcolsep 10pt
\begin{tabular}{l c c  c c}  \hline  
  &N.of.S & $\gamma$  &  $x_0$  & $x_0$ ($\gamma=1.8$)  \\  \hline
$w_p(r_p)$ & 186 & 1.7$\pm0.1$ & 4.4$\pm0.7$ & $4.5\pm0.7$ \\
$\xi(r_p)$ &     & 1.4$\pm0.1$ & 6.3$\pm1.5$ & $5.4\pm1.0$ \\ \hline
\end{tabular} 
\end{table}

\begin{figure}
\includegraphics[
width=8.1cm]{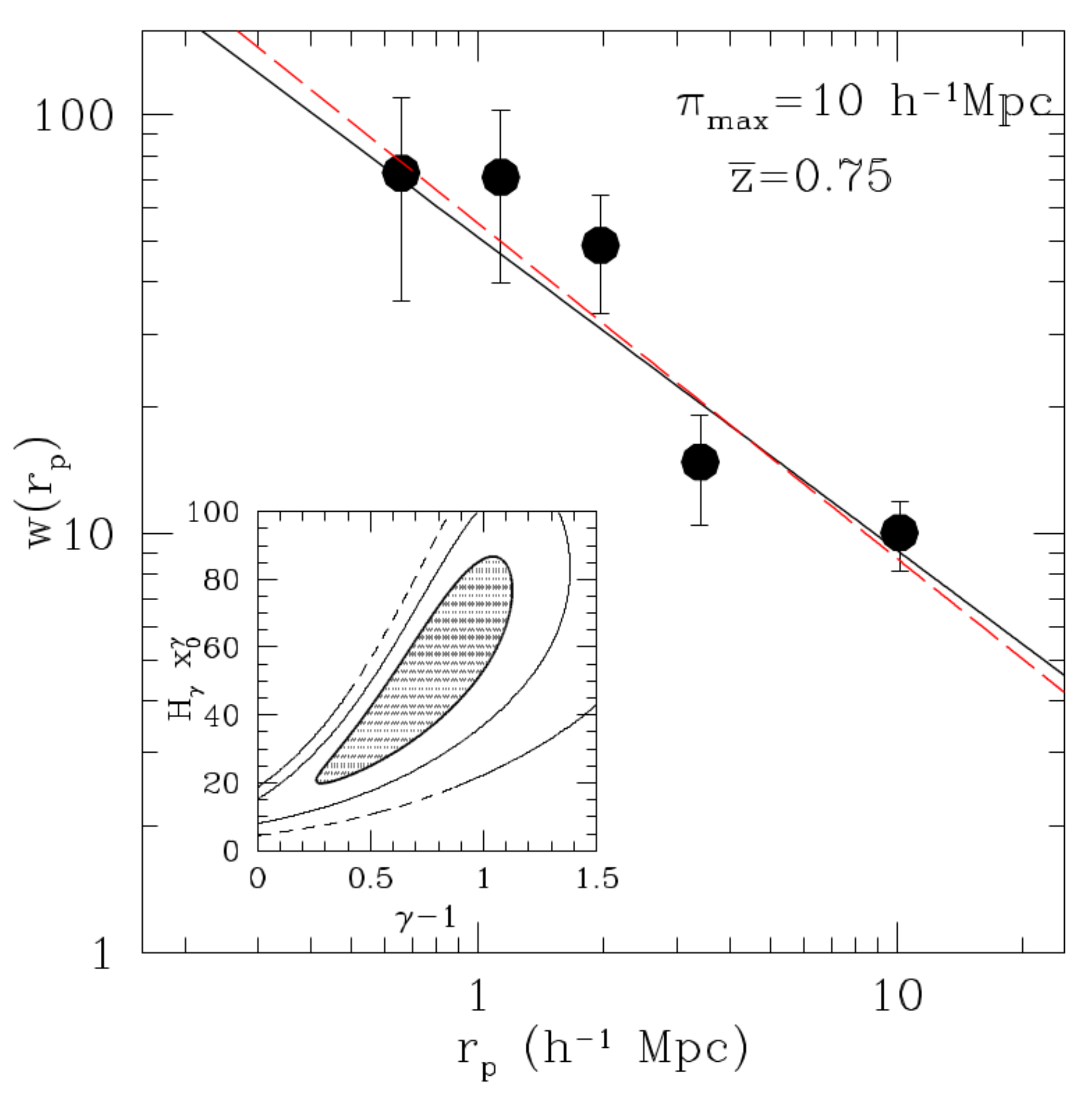} 

\caption{The projected $w_p(r_p)$ correlation function for the AEGIS
  field. The black line corresponds to the fit with free $\gamma$, while the
  red line to that for $\gamma=1.8$. The inset panel show the 1, 2 and
  3$\sigma$ likelihood contours in the 2-parameter plane of power law
  solutions. Note that the y-axis of the inset plot is the combined parameter
  $H_{\gamma} x_0^\gamma$, since $H_\gamma$ also depends on the free
  parameter $\gamma$ (inspect eq.\ref{eq:wp_model}).}
\end{figure}

\section{RESULTS}
\subsection {Direct Spatial Correlation Function}
In order to estimate the projected correlation function, $w_p(r_p)$,
we use the estimator
provided by eq.(\ref{LS}) and 12 logarithmic separation bins covering the range
$0.5<r_p<40 h^{-1}$ Mpc.
As for the choice of $\pi_{\rm max}$, it should be a compromise between having an
optimal correlation signal to noise ratio and reducing
the excess noise from high $\pi$ separations, which are affected by
redshift-space distortions.
We have investigated the sensitivity of $w_p(r_p)$ on $\pi_{\rm max}$, which
we have varied in 
the range $[5, 25] h^{-1}$ Mpc; \citep[see also][]{Koutoulidis13} and
found that it is quite stable. We present the results
based on $\pi_{\rm max}=10 h^{-1}$ Mpc.

In Figure 3 we present the derived hard-band projected correlation function. 
The results of the corresponding power-law fits to the correlation function data
are listed in Table 1.

Using the G05 method to construct the random catalogue, 
we find the best fit correlation length of $w_p(r_p)$ 
to be $x_0=4.4\pm 0.7 \; h^{-1}$ Mpc (with $\gamma=1.7\pm0.1$), which is in
excellent agreement with the total (0.5-8 keV) band result
($x_0=4.3\pm0.6 \;h^{-1}$ Mpc, $\gamma=1.6\pm0.1$, reported in Table 2
of Koutoulidis et al. 2013).
Once we correct for the factor $C_\gamma(r_p)$ (eq. \ref{eq:CorF}),
the above result translates to 
a $\xi(r_p)$ with $x_{0,c}= 5.4 \pm 1.0 \; h^{-1}$ Mpc (for
$\gamma=1.8$), which implies that had we not corrected for the
instrinsic underestimation of $\xi(r)$ when using the $w_p(r_p)$
estimator and a finite value of $\pi_{max}$ (in our case 10$h^{-1}$
Mpc), we would have underestimated the true
correlation length by $\sim 20\%$. 

We have also tested and found that these
results remain robust when changing the random construction method to
that based on the sensitivity map.
Our results also agree with other previously derived clustering
results of the same field.
 \citet{Coil09} derived the AGN/galaxy cross-correlation using 113
 Chandra AGN in the full 0.5-7 keV and found $x_0=5.9 \pm 0.9
 h^{-1}$Mpc at a median $\bar{z}\simeq 0.9$.
 
\subsection {Angular Correlation function}
In Table 2 we present the best fit values of the ACF parameters $\gamma$ and $\theta_0$ 
for the whole and for the spectroscopic subsample, while in Figure 4 we plot 
the corresponding $w(\theta)$ for both samples. It is evident that the
angular clustering amplitude for the spectroscopic sample 
({\em Spec}) appears significantly larger than that of the whole
sample ({\em All}). We should also point out that the $w(\theta)$ of
the {\em All} sample has negative values at some separations (as can
be inferred from Fig.4), while the power-law fit 
was performed only on the positive values, a fact which implies that the
derived amplitude is an upper limit to the true clustering of this sample.
The apparent difference between the ACF of the {\em Spec} and {\em
  All} samples could be attributed to the dependence of clustering on
the limiting flux \citep[e.g.,][]{Plionis08,Ebrero09,Elyiv12}. Indeed,
as we show in Figure 5, the flux distributions for the spectroscopic
sample (red shaded region) is shifted to higher fluxes with respect to
the complete sample of sources ({\em All}, black thick
line), with corresponding mean fluxes are $f_x \simeq 7.8\times 10^{-15}$
and $5.4\times 10^{-15}$ erg s$^{-1}$ cm$^{-2}$, respectively. The
observed difference is indeed statistically significant as shown by the
Kolmogorov-Smirnov two-sample test which provides a probability of consistency
between the two distributions of only $\sim 2\times 10^{-8}$.
\begin{figure}
\begin{center}
 \includegraphics[
width=8.6cm]{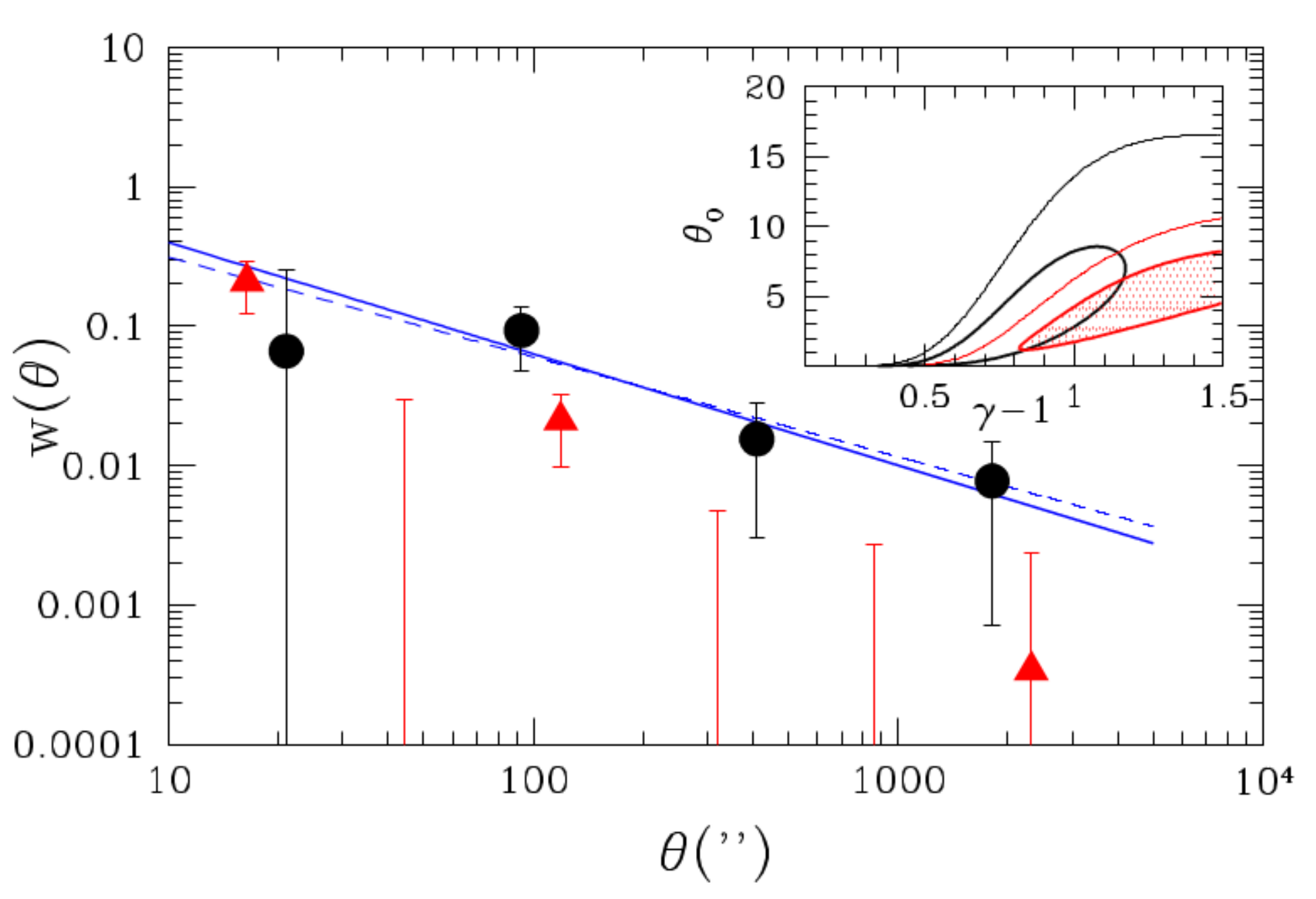} 
\end{center}
\caption{The angular correlation function of the AEGIS field for the subsample with spectroscopic 
information within $0.3<z<1.3$ (i.e 186 sources, black filled circles) and for the
complete X-ray source  sample (i.e 741 sources, red filled triangles)
in the hard band. The solid line corresponds to the fixed 
$\gamma=1.8$ fit, while the dashed line represents the best power-law fit. The error bars 
corresponds to 1$\sigma$ uncertainties. The inset plot presents the 1 and 3$\sigma$ contours in 
the fitted $(\theta_0, \gamma)$ parameter space.}
\end{figure}

\begin{table}
\caption{AEGIS hard-band angular correlation function results for the
  complete sample ({\em All}) and for those sources with 
spectroscopic redshifts ({\em Spec}, $0.3<z<1.3$).}
\label{tab_sample}
\tabcolsep 12pt
\begin{tabular}{l c c c c} \hline  

 & $N$ & $\gamma$  &  $\theta_0/^{''}$  & $\theta_0/^{''}$ ($\gamma=1.8$)  \\  \hline 
{\em All} & 741 & 2.1$\pm 0.2$ & 4.3$\pm1.2$ & $ 1.2\pm0.5$ \\
{\em Spec} &186 & 1.7$\pm 0.1$ & 1.6$\pm1.0$ & $ 2.9\pm1.4$ \\
\hline
\end{tabular} 
\end{table}

\begin{figure}
\includegraphics
[width=8.1cm]{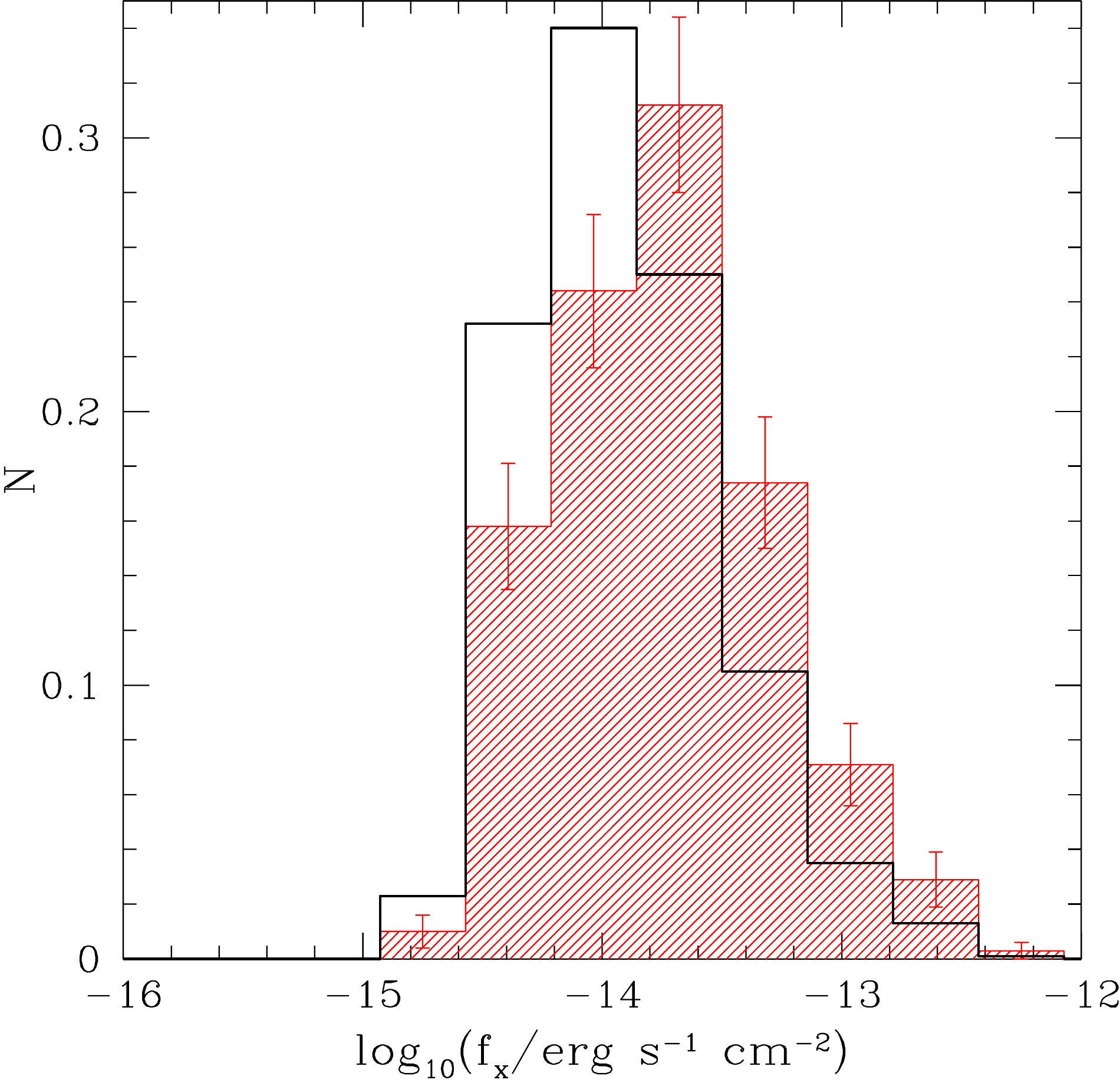} 
\caption{Comparison of the normalized flux distributions of the
  complete sample (thick black line) and of the
  spectroscopic subsample (red shaded region).}
\end{figure}

\subsection{Correlation Length by inverting $w(\theta)$}
In order to derive the spatial clustering length from the angular
correlation function we use eq.(\ref{eq:inv22}) and as the source
$N(z)$ distribution we use either the integral of the hard-band 
Luminosity and Density Evolution (LADE) luminosity function \citep{Aird2010}
or directly the redshift distribution of the sources with available
spectroscopic redshifts.
Both approaches lead to exactly the same results and thus 
we present results based on the later approach.

The next step in order to derive the spatial clustering length 
is to somehow estimate the exponent of the growth factor, ${\tilde
  D}^{2+n}(z)$. One could 
therefore use literature clustering estimates of 
hard-band AGN samples, dominating different redshits, and fit eq.(13)
to such data. Since however, the available hard-band
results are very few; ie., that of \citep{Mountrichas2012} with 
$r_0=4.8 \pm 1 \;h^{-1}$ Mpc at $\bar{z}=0.1$ and of CDF-N with 
$r_0=5 \pm 1 h^{-1}$ Mpc at $\bar{z}=0.9$\citep{Gilli05}
\footnote{The  correlation length of CDF-S ($r_0=9.8 \pm
  1 h^{-1}$Mpc at $\bar{z}=0.7$) is not considered due to the presence
  of superclusters which affect the clustering amplitude
  \citep{Gilli03}.}, we will also use total band (0.5-8 keV) clustering results
 \citep[e.g.,][]{Coil09, Starikova11, Allevato11, Koutoulidis13}
excluding the results of $\bar{z}\simeq 0.75$ which appears to present
an erratic behaviour in all available studies (see Fig.8
and relevant discussion in Koutoulidis et al. 2013). 
It appears that there is a weak but
consistent increase of the value of $x_0$ with redshift.
The use of total band results is also supported by the fact that the
corresponding AEGIS correlation function results are in excellent agreement
with those of the hard-band, as discussed previously.

Subsequently, we perform a $\chi^2$ minimisation fitting of the above
data to eq. (13), leaving as a free parameter the exponent $n$ and fixing the halo
mass to $M_h=1.3 \times 10^{13} M_{\odot}$, which we derived in \cite{Koutoulidis13}. 
The resulting best fit value for the exponent is $2+n=5.03\pm 0.2$. 
The corresponding $\chi^2-\chi^2_{min}$ and clustering evolution
curves can be seen in the left and right panels of Figure 6,
respectively (in black). 
Alternatively we can leave as free parameters both the exponent $n$
and the halo mass $M_h$. In this
case the minimization procedure provides a degenerate solution,
similar to that of Fig.2. As an example, we present in the right panel
of Figure 6  another clustering evolution model (red curve), that
with $(n, M_n)\simeq (2.0, 2\times 10^{12} M_{\odot})$, consistent
with the data and selected from within the 1$\sigma$ range of the
degenerate solution.

Taking into account the above parameters and using eq.(\ref{eq:inv22})
 we deproject the ACF and derive the spatial clustering length.
Since the Limber's inversion provides the spatial clustering
length at $z=0$, in order to predict its value at another redshift it
is necessary to normalized eq.(13) to that redshift.
As seen in Table 3, normalizing to $z=0.75$, which corresponds to the median redshift of 
the spectroscopic AEGIS hard-band sample, we find an excellent agreement
between the direct measure of $x_0$ and the corresponding Limber's
inverted one. Similarly, normalizing to $z=0.1$ which is
the median redshift of the sample studied in \citep{Mountrichas2012}
we again find an excellent agreement with the direct $x_0$ measure 
(see Table 3).

\begin{figure*}
\begin{center}
\includegraphics
[width=16cm]{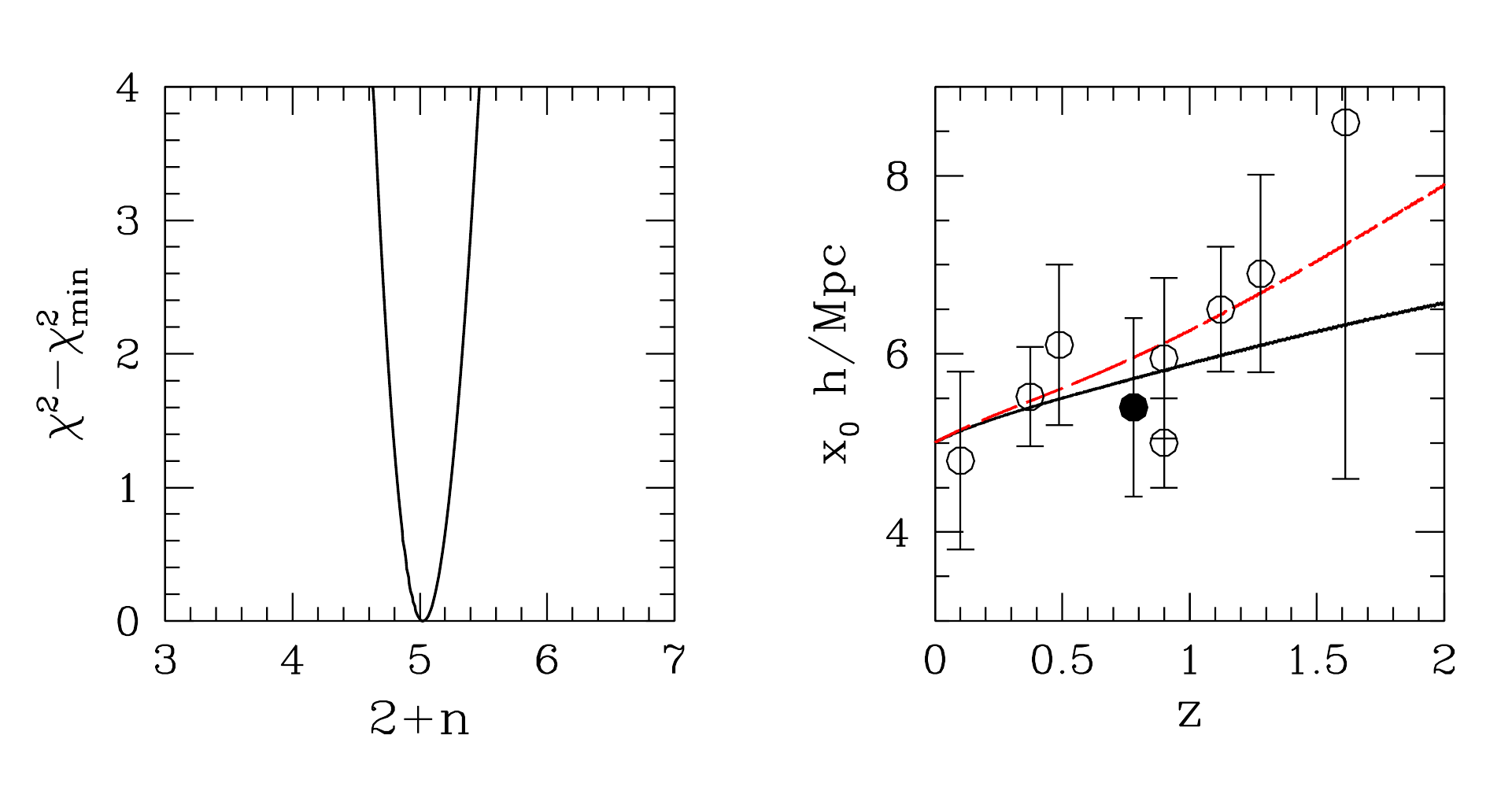} 
\end{center}
\caption{{\em Left panel}: Result of the $\chi^2$ minimisation
  procedure of fitting the parameter $2+n$.
{\em Right panel}: The clustering evolution model (black curve),
corresponding to the best fit value ($n=3.03$) for
$M_h=1.3\times10^{13} M_{\odot}$, overplotting the literature $x_0$ data 
in the hard and total band. Our direct estimate (not used in the $\chi^2$
minimization procedure) is shown as the filled circle. 
As a manifestation of the degeneracy discussed in section 2, we show
the expected clustering evolution for another $(n, M_n)$ pair 
$\simeq (2.0, 2\times 10^{12} M_{\odot})$
which is consistent with the data  (red curve).}
\end{figure*}

\begin{table}
\caption{Comparison of the inverted ACF AEGIS hard-band correlation
  length, $x_0$ (in $h^{-1}$ Mpc), and that directly measured by at
  $\bar{z}=0.1$ (Moutrichas et al. 2012) and $\bar{z}=0.75$ (our
  current measure).}
\tabcolsep 28pt
\begin{tabular}{c c c } 
\hline  
$\bar{z}$ & $w(\theta)$ based & $\xi(r_p)$ based\\  \hline 
0.10       & 5.0$\pm 1.1$       &$4.8\pm 1.0$  \\
0.75       & 5.5$\pm 1.2$       &$5.4\pm 1.0$  \\
 \hline
\end{tabular} 
\end{table}

Since the parameterisation of X-ray AGN clustering evolution (eq.\ref{eq:xi-r})
reproduces the direct spatial correlation length for a particular
value of $n$, we apply it also to the clustering results of the whole
AEGIS X-ray source sample ($\theta_0=1.2 \pm 0.6$ arcsec with $\gamma=1.8$ -
Table 2). 
We use the expected $N(z)$, based on the LADE luminosity function
of \cite{Aird2010}, but for the whole interval $z=0-4$, since we
cannot a priori assess if the sources without 
redshifts correspond to nearby low-luminosity or distant high-luminosity
sources. The predicted median redshift of this sample is $\bar{z}=0.98$.
Using Limber's inversion we obtain $r_0=4.8 \pm 0.9\; h^{-1}$ Mpc at
$z=0.1$ and $r_0=5.5 \pm 1.0\; h^{-1}$ Mpc at $z=0.1$ and $z=0.75$,
respectively, which are in excellent agreement with the
scaled $0.3<z<1.3$ and the direct $\xi(r_p)$ results (see Table 3).
Given these results we can infer that the sources without redshift information 
seem to follow roughly the same clustering evolution as the observed.
Moreover the above procedure indicates the 
potential for an accurate derivation of the spatial correlation length
 using only the angular clustering of all the available sources,
 independently of availability or not of complete spectral
 information, for as long as the luminosity function in hand relates
 to the whole sample of sources.

\section{DISCUSSION \& CONCLUSIONS}
A variety of studies have measured the projected angular clustering
of X-ray AGN, and then via Limber's equation derived
the corresponding spatial clustering length. Although there are
obvious merits in this approach, ie., the fact that one uses all the
available sources while being unaffected by redshift-space
distortions, there are disadvantages and inherent assumptions in the
deprojection the most important of which is the unkown clustering
evolution model of the sources. 
There have been interesting parametrizations of such an evolution,
the most common of which is the $1+z$ power law model (the
so-called $\epsilon$ models; see Eq.\ref{xi223}), which has been used
extensively to model the evolution of clustering. However, it is
rather phenomenological in nature and thus it does not provide a good
physical insight into the evolution governed by the clustering of dark
matter halos \citep{McCracken01}.
In this work we presented a more generic parametrization that
allows for different cosmological models and different host dark
matter halo mass, via their different bias evolution, and we showed the
regimes where the two parametrizations coincide.

One can identify two interesting extremes of the clustering evolution; 
The stable clustering scenario ($\epsilon=0$ or $n\gtrsim 4.2$
depending also on the halo mass, for the
two different parametrizations, respectively) in which clusters remain
bound and stable while due to the the expansion the background density
drops by $(1+z)^3$
and the constant in comoving coordinates scenario ($\epsilon=\gamma-3$
or $n\lesssim 3.5$; see section 2) in which pairs of sources follow
the Hubble flow, meaning that their separation
remains constant in comoving coordinates. 
Interestingly, the correlation function analyses of optically selected
QSO samples\citep{Croom05,Ross09} seem to suggest a roughly 
comoving evolution model of their clustering, while simulations also seem to
disfavour the stable clustering evolution model\citep{Jain97}.

In this work, we used the comoving clustering lengths of 
X-ray AGN at a variety of different redshifts, 
provided in the literature 
\citep{Gilli05, Coil09, Starikova11, Allevato11, Mountrichas2012,
  Koutoulidis13},  to fit the unknown slope of the
non-linear contribution to
the growing mode of perturbations in order to model the evolution of
clustering of X-ray AGN. We then applied this to the Limber's
inversion of the angular correlation function of the subsample of 
AEGIS X-ray AGN
with spectroscopic data within $0.3<z<1.3$ to find an excellent
agreement with the directly measured spatial clustering of the same
sources.

 The reason of the discrepancies between angular and spatial clustering
results of various previous studies is still not clearly identified,
although the clustering evolution parametrization, used in this work,
does hint towards the possibility that the different host halo mass for the
different sources, and thus their different bias evolution, induce a
different clustering evolution than what one assumes when using the
phenomenological $\epsilon$ parametrization.

The present work appears to pave the way for the application of the
angular correlation function analysis on large X-ray selected
AGN samples, such as the eROSITA survey which is expected to detect
over $3\times 10^6$ AGN. Such numbers of sources render 
rather impossible the use of spectroscopic redshifts to derive their
spatial clustering, as this would demand unrealistically large
follow-up optical telescope time.

\bibliography{ref}{}
\bibliographystyle{aa}

\end{document}